# Leveraging Artificial Intelligence to Promote Awareness in Augmented Reality Systems

WANGFAN LI, Clemson University, USA
ROHIT MALLICK, Clemson University, USA
CARLOS TOXTLI-HERNANDEZ, Clemson University, USA
CHRISTOPHER FLATHMANN, Clemson University, USA
NATHAN J. MCNEESE, Clemson University, USA

## 1 POSITION STATEMENT

Recent developments in artificial intelligence (AI) have permeated through an array of different immersive environments, including virtual, augmented, and mixed realities. AI brings a wealth of potential that centers on its ability to critically analyze environments, identify relevant artifacts to a goal or action, and then autonomously execute decision-making strategies to optimize the reward-to-risk ratio. However, the inherent benefits of AI are not without disadvantages as the autonomy and communication methodology can interfere with the human's awareness of their environment. More specifically in the case of autonomy, the relevant human-computer interaction literature cites that high autonomy results in an "out-of-the-loop" experience for the human such that they are not aware of critical artifacts or situational changes that require their attention. At the same time, low autonomy of an AI system can limit the human's own autonomy with repeated requests to approve its decisions. In these circumstances, humans enter into supervisor roles, which tend to increase their workload and, therefore, decrease their awareness in a multitude of ways. In this position statement, we call for the development of human-centered AI in immersive environments to sustain and promote awareness. It is our position then that we believe with the inherent risk presented in both AI and AR/VR systems, we need to examine the interaction between them when we integrate the two to create a new system for any unforeseen risks, and that it is crucial to do so because of its practical application in many high-risk environments.

With the recent advancement in Artificial intelligence(AI) in the form of Large Language Model(LLM) and Large Multimodel Model(LMM), such as ChatGPT, and advancement in Augmented Reality(AR) and Virtual Reality(VR )devices, such as Apple Vision Pro and Meta Quest 3, we can expect that further development will come to both the field of AI and AR, so will the adoption of their usage. Although both AI and AR offer unparalleled innovation and opportunity, they also come with plenty of safety concerns that, because of the rate of development, are still being very actively researched. In the case of LLM and LMM, some of the risks are known and documented; for example, we know that they are susceptible to both prompt injection that makes the model produce unintended outputs[5, 9] despite safety measures and also their potential for mass producing misinformation online[2, 3]. On the AR and VR side of things, the question of safety is becoming more critical as it integrates more into everyday consumer use and application, not just its usage as a social tool, in which many risks can cause harm[1, 12], and also in its usage as a practical work improvement tool in essential areas such as construction that have a low tolerance for error and risks[6]. There are some perceived benefits in both AI and AR, so it makes sense first to combine the two into a system to aid the user in various tasks by combining the immersive nature of AR and the capabilities of AI[11]; however, the usage risks that arise when we combine AR and AI systems into one application are not fully explored.

We are a group of researchers working broadly in the human-computer interaction field with diverse levels of experience and backgrounds. With years of research experience collectively, we are concerned not only with advancing the field of immersive virtual space innovation-wise but also with scrutinizing their ethical and safety implications. We are also currently collaborating on a research project that we feel fits the workshop's theme well, providing a good opportunity for us to contribute.

This is the primary motivation for the research project we are currently collaborating on, and it is about examining various factors of the design of a combined AR and AI system, specifically when and how the AR and AI elements engage the user and their impact on different types of user awareness[7]. The factors we are examining here are explainability[4], representing how much information the AI provides to the user in an AR environment, and autonomy[8], representing how much control the user has over the process of the AI system. From previous research, we know that interface design can have a large impact on awareness, especially in the case of complex interfaces that cause information overload[10], and the two factors here can both increase the complexity of the system, while also interacting with each other and the AR setting when it comes to design, the resulting impact on awareness is unclear. Awareness is critical for user safety while using the system, as a badly design interface could directly reduce user awareness and lead to physical harm in the real world, especially in a complex environment such as a warehouse or factory.

We hope that by participating in the workshop, we can listen and collaborate with other researchers regarding risk and safety in the emerging immersive reality technology. By bringing our unique perspective and experience regarding immersive reality technology and our current collaboration on AI and AR integration both as a new avenue of risk and a way to reduce risk, it is a chance to share what we are currently working on and also to get feedback and collaboration in this field, with the end goal being a safer immersive virtual space.